\documentstyle[preprint,tighten,aps,epsfig]{revtex}
\begin{document}
\title{Faddeev study of heavy baryon spectroscopy}
\author{H. Garcilazo$^{(1)}$, J. Vijande$^{(2,3)}$, A. Valcarce$^{(3)}$} 
\address{$(1)$ Escuela Superior de F\'\i sica y Matem\'aticas, \\
Instituto Polit\'ecnico Nacional, Edificio 9,
07738 M\'exico D.F., Mexico}
\address{$(2)$ Dpto. de F\' \i sica Te\'orica and IFIC,
Universidad de Valencia - CSIC,\\
E-46100 Burjassot, Valencia, Spain}
\address{$(3)$ Departamento de F{\'\i}sica Fundamental, \\
Universidad de Salamanca, E-37008 Salamanca, Spain}
\maketitle

\begin{abstract}
We investigate the structure of heavy baryons containing
a charm or a bottom quark. We employ a constituent 
quark model successful in the description of the 
baryon-baryon interaction which is consistent 
with the light baryon spectra. 
We solve exactly the three-quark 
problem by means of the Faddeev method in momentum space. 
Heavy baryon spectrum shows a manifest 
compromise between perturbative and
nonperturbative contributions. The 
flavor dependence of the one-gluon exchange is analyzed.
We assign quantum numbers to some already observed
resonances and we predict the first radial and orbital excitations
of all states with $J=1/2$ or $3/2$.
We combine our results with heavy quark symmetry
and lowest-order $SU(3)$ symmetry breaking to predict the masses
and quantum numbers of six still non-measured ground-state beauty baryons.
\end{abstract}

\vspace*{2cm} \noindent Keywords: \newline
nonrelativistic quark models, baryon spectra
\newline
\newline
\noindent Pacs: \newline
12.39.Jh, 12.39.Pn, 14.20.-c

\newpage

\section{Introduction}

Since the hypothesis at BNL \cite{Caz75} and posterior confirmation 
at Fermilab \cite{Bal79} of the existence of charmed baryons, an increasing
interest on heavy baryon spectroscopy arose. It became evident that baryons
containing heavy flavors $c$ or $b$ could play an important role in our
understanding of QCD. In the early 90's the first beauty
baryon, $\Lambda_b$, was discovered at the CERN $e^+ e^-$ collider 
LEP \cite{Alb91}. Since then, several new hadrons containing a single 
charm or bottom quark have been identified \cite{Eid04}, they are 
resumed in Table \ref{t1}. Recent years have been specially fruitful 
in discovering heavy baryon states \cite{Miz05,Aub06,Fei95,Edw95,Art01},
and new states will be soon reported \cite{Les06}.
In many cases signals for the production of charmed
and beauty baryons have been observed, the definitive measurement
of the mass awaiting confirmation. This is, for example, the
case of $\Xi_b$ and $\Omega_b$ baryons already produced in
$Z^0 \to b\overline b$ decays \cite{Bus96}. Thus, there are
more states still to be found \cite{Ros06}.

While the mass of these particles is usually measured 
as part of the discovery process, other quantum numbers such as the spin or
parity often prove more elusive. For heavy baryons,
no spin or parity quantum numbers of a given state have been measured directly. 
These properties can only be extracted by
studying angular distributions of the particle decays, but these
are available only for the lightest and most abundant species.
For excited heavy baryons the data set
are typically one order of magnitude smaller than for heavy mesons. 
Besides, knowledge of orbitally excited states is very much limited.
Therefore, a powerful guideline for assigning quantum numbers 
to new states or even to indicate new states to look for is 
required by experiment. 

All these reasons make heavy baryon spectroscopy an extremely rich 
and interesting subject. On the one hand
it represents a three-body problem that may be solved exactly within
the Faddeev formalism. On the other hand, it allows to study how the
interaction between quarks evolves from light to 
heavy systems. Understanding baryon spectroscopy is expected to
improve our knowledge about basic properties of QCD. In particular,
heavy baryons provide with an excellent and perhaps even dramatic way
of testing the approximate flavor independence of confinement forces.
Besides, heavy baryons provide an excellent laboratory to study the
dynamics of a light diquark in the environment of a heavy quark, allowing the
predictions of different theoretical approaches to be tested. 

After the discovery of the first charmed baryons, several theoretical works 
\cite{Cop79,Mal80,Ric83} based on potential models developed 
for the light baryon or meson spectra started analyzing properties of
the observed and expected states. Later on, charmed and beauty baryons were studied
with different Bhaduri-like potentials \cite{Sil96}, calculating the ground
state by a Faddeev method in configuration space and the excited states
by diagonalization in a harmonic oscillator basis up to eight quanta. 
The charmed baryon spectrum has also been recently analyzed by means of 
a relativistically covariant quark model based on the
Bethe-Salpeter equation in the instantaneous approximation \cite{Mig06}.
Finally, the ground states of charmed and beauty baryons have been 
calculated by means of lattice techniques \cite{Bow96}. 

Nowadays, we have to our disposal {\it realistic} quark models accounting
for most part of the one- and two-body low-energy hadron phenomenology.
Among the several quark models proposed in the
literature \cite{Val05}, either they were designed
to study the baryon-baryon interaction 
\cite{Shi89,Bra90,Wan92,Yuz95,Fuj04} 
or the baryon spectra 
\cite{Isg79,Sil85,Des92,Dzi96,Glo96,Fur02}.
To our knowledge, the ambitious project of a simultaneous description 
of the baryon-baryon interaction and the baryon (and meson) 
spectra has only been undertaken by the constituent quark 
model of Ref. \cite{Val05}. It was originally designed 
to study the nonstrange sector and it has been recently generalized
to all flavors \cite{Vij04}. The success in describing the 
properties of the strange and non-strange one and two-hadron
systems encourages its use as a guideline in order to assign parity 
and spin quantum numbers to already determined heavy baryon
states as well as to predict still non-observed resonances. 

This is why in this work we pursue the study of heavy baryons 
containing a charm or a bottom quark making use of the constituent 
quark model of Ref. \cite{Vij04}. Consistency with the baryon-baryon
interaction and the light-baryon spectrum will be required.
For the first time a Faddeev calculation
in momentum space is done for the ground and excited states of baryons made
of a heavy $c$ or $b$ quark.
This is also the first time we study systems made
of three distinguishable particles.  
The paper is organized as follows, 
we will start in the next section resuming the basic properties
of the constituent quark model and describing the
Faddeev method in momentum space for three
distinguishable particles.
Section \ref{sec4} will be devoted to present 
and discuss the results in comparison to other models in the literature. 
Finally, in Sec. \ref{sec5} we will summarize our conclusions.

\section{Working framework}
\label{sec2}

\subsection{Constituent quark model}

Let us first outline the basic ingredients of the constituent quark model of
Ref. \cite{Vij04}.
Since the origin of the quark model hadrons have been considered to be
built by constituent (massive) quarks. Nowadays it is widely
recognized that the constituent quark mass appears because of the spontaneous
breaking of the original chiral symmetry of the QCD Lagrangian, which 
gives rise to boson-exchange interactions between quarks ($V_{\chi}$).

QCD perturbative effects are taken into account
through the one-gluon-exchange (OGE) potential \cite{Ruj75}.
The $\delta$-function, arising as a consequence of the nonrelativistic
reduction of the one-gluon exchange diagram between point-like
particles, has to be regularized in order to perform exact calculations.
This regularization, controlled by a parameter $r_0$, has to be
flavor dependent \cite{Wei83}. As this will be a central issue
of our discussion, we will make explicit the expressions concerning
the OGE. It reads,  
\begin{equation}
V_{OGE}(\vec{r}_{ij}) ={\frac{1}{4}}\alpha _{s}\,\vec{\lambda ^{c}}%
_{i}\cdot \vec{\lambda^{c}}_{j}\,\left\{ {\frac{1}{r_{ij}}}-{\frac{1}{%
6m_{i}m_{j}}}\vec{\sigma}_{i}\cdot \vec{\sigma}_{j}
\,{\frac{{e^{-r_{ij}/r_{0}}}}{r_{ij}\,
r_0^2}}\right\} \, ,
\end{equation}
where $\lambda^{c}$ are the $SU(3)$ color matrices, 
$\alpha_s$ is the quark-gluon coupling constant, and
$r_0$ is a flavor-dependent parameter to be determined from data.
The scale-dependent strong coupling constant \cite{God85}
is given by \cite{Vij04}, 
\begin{equation}
\alpha_s(\mu)={\alpha_0\over{ln\left[{({\mu^2+\mu^2_0})/
\gamma_0^2}\right]}},
\label{asf}
\end{equation}
where $\mu$ is the reduced mass of the interacting $qq$ 
pair and $\alpha_0=2.118$, 
$\mu_0=36.976$ MeV and $\gamma_0=0.113$ fm$^{-1}$.

Finally, any model imitating QCD should incorporate
confinement (CON). Lattice calculations
in the quenched approximation derived, for heavy quarks, a
confining interaction linearly dependent on the interquark
distance.  The consideration of sea quarks apart from valence 
quarks (unquenched approximation) suggests a screening effect 
on the potential when increasing the interquark distance \cite{Bal01}.

Once perturbative (one-gluon exchange) and nonperturbative (confinement and
chiral symmetry breaking) aspects of QCD have been considered, one ends up with
a quark-quark interaction of the form 
\begin{equation} 
V_{q_iq_j}=\left\{ \begin{array}{ll} 
q_iq_j=nn/sn\Rightarrow V_{CON}+V_{OGE}+V_{\chi} &  \\ 
q_iq_j=cn/cs/bn/bs/cc/bb \Rightarrow V_{CON}+V_{OGE} & 
\end{array} \right.\,.
\label{pot}
\end{equation}
Notice that for the particular case of heavy quarks ($c$ or $b$) chiral symmetry is
explicitly broken and therefore boson exchanges do not contribute.
For the explicit expressions of the interacting potential and a more
detailed discussion of the model we refer the reader to Refs. \cite{Vij04,Val05b}.
For the sake of completeness we resume the parameters of the model in Table \ref{t2}.

We have not considered the noncentral contributions arising from
the different terms of the interacting potential. Experimentally, there
is no evidence for important effects of the noncentral terms on the
light baryon spectra. This is clearly observed in the almost degeneracy of the
nucleon ground states with $J^P=1/2^-$ and $J^P=3/2^-$, or their
first excited states with the nucleon ground state with $J^P=5/2^-$.
The same is observed around the whole
baryon spectra except for the particular problem of the
relative large separation between the $\Lambda(1405)$, $J^P=1/2^-$,
and the $\Lambda(1520)$, $J^P=3/2^-$, related to the
vicinity of the $N \overline K$ threshold \cite{Vei85}.

Theoretically, the
spin-orbit force generated by the OGE has been justified to cancel with
the Thomas precession term obtained from the confining potential \cite{Isg00}.
This is not however the case for the two-baryon system where,
by means of an explicit model for confinement, it has been demonstrated
that the strong cancellation in the baryon
spectra translates into a constructive effect
for the two-baryon system \cite{Koi86}.
One should notice that the scalar boson-exchange potential also
presents a spin-orbit contribution with the same properties as before,
it cancels the OGE spin-orbit force in the baryon spectra
while it adds to the OGE contribution for the nucleon-nucleon $P-$waves
and cancels for $D-$waves \cite{Val95},
as it is observed experimentally.
Such a different behavior in the one- and two-baryon systems
is due to the absence of a direct term in the OGE spin-orbit force
(due to the color of the gluon only quark-exchange
diagrams are allowed), while the spin-orbit contribution of the
confining interaction in Ref. \cite{Koi86} and that of the
scalar boson-exchange potential in Ref. \cite{Val95}
are dominated by a direct term, without
quark exchanges. Regarding
the tensor terms of the meson-exchange potentials, they have been explicitly
evaluated in the light-baryon case (in a model with stronger meson-exchange
potentials) finding contributions not bigger that 25 MeV \cite{Fur03}.
This is due to the fact that the tensor terms give
their most important contributions at intermediate
distances (of the order of 1$-$2 fm), due to the direct term in the
quark-quark potential. The regularization of the boson-exchange
potentials below the chiral symmetry breaking scale
suppresses their contributions for the very small distances
involved in the one-baryon problem.
This allows to
neglect the noncentral terms of the interacting potential
that would provide with a fine tune of the final
results and would make very much involved and time-consuming
the solution of the three-body problem by means of the Faddeev method
in momentum space we pretend to use.
The same result has been obtained in Ref. \cite{Mig06} in the
study of the noncentral terms of instanton induced interactions
for charmed baryons.

The noncentral terms are relativistic corrections whose effect is
known to decrease for heavy baryons \cite{Mig06,Ped03},
either due to the absence of the interaction (boson exchanges) or to the fact
that they are $1/m^2$ corrections (OGE). In particular,
the noncentral terms have also been explicitly evaluated
in the past by the present authors, demonstrating that their contribution
is already very small in the light baryon sector.
The tensor potential of the
one-pion exchange potential has been explicitly evaluated in Ref. \cite{Val96},
obtaining contributions smaller than 20 MeV. Spin-orbit terms
have been calculated in Ref. \cite{Gar05} by means
of a relativistic treatment of the spin variables, obtaining
corrections one order of magnitude smaller than those due to the
use of relativistic momentum variables.

\subsection{Three-body equations}

After partial-wave decomposition, the Faddeev equations are 
integral equations in two continuous variables 
as shown in Ref. \cite{Val05b}. They can be transformed into integral
equations in a single continuous variable 
by expanding the two-body
t-matrices in terms of Legendre polynomials as shown in Eqs. (32)$-$(36)
of Ref. \cite{Ter06}.
One obtains the final set of equations

\begin{eqnarray}
\psi_{i;LST}^{n\ell_i\lambda_iS_iT_i}(q_i) = \sum_{j\ne i}
\sum_{m\ell_j\lambda_jS_jT_j}\int_0^\infty q_j^2 dq_j \,
K_{ij;LST}^{n\ell_i\lambda_iS_{i}T_{i}m\ell_j\lambda_jS_{j}T_{j}}(q_{i},q_{j};E)
 \nonumber \\
 \times 
\psi_{j;LST}^{m\ell_j\lambda_jS_jT_j}(q_j),
\label{e22c5}
\end{eqnarray}
with

\begin{eqnarray}
K_{ij;LST}^{n\ell_i\lambda_iS_{i}T_{i}m\ell_j\lambda_jS_{j}T_{j}}(q_{i},q_{j};E)
&=& {1\over 2}
 <S_iT_i|S_jT_j>_{ST}\
\sum_{r}\tau
_{i;nr}^{\ell_iS_{i}T_{i}}(E-q_{i}^{2}/2\nu _{i})  \nonumber \\
&&\times \int_{-1}^{1}d{\rm cos}\theta \,{\frac{P_{r}(x^\prime_{i})P_{m}(x_{j})}{%
E-p_{j}^{2}/2\eta _{j}-q_{j}^{2}/2\nu _{j}}}
A_L^{\ell_i\lambda_i\ell_j\lambda_j}(p_i^\prime q_i p_j q_j).
\label{e23c5}
\end{eqnarray}
$P_{r}(x^\prime_{i})$ and $P_{m}(x_{j})$ are Legendre polynimials,
$x_i^\prime=(p_i^\prime-b)/(p_i^\prime+b)$,  
$x_j=(p_j-b)/(p_j+b)$, and $b$ a scale parameter.
$\tau_{i;nr}^{\ell_iS_{i}T_{i}}(E-q_{i}^{2}/2\nu _{i})$ are the
coefficients of the expansion of the two-body $t-$matrices in terms
of Legendre polynomials defined
by Eq. (34) of Ref. \cite{Ter06}.
$S_i$ and $T_i$ are the spin and isospin of the 
pair $jk$ while $S$ and $T$ are the total spin and isospin.
$\ell_i$ is the orbital angular momentum 
of the pair $jk$, $\lambda_i$ 
is the orbital angular momentum 
of particle $i$ with respect to the pair
$jk$, and $L$ is the total orbital angular momentum. 
The reduced masses $\eta_i$ and $\nu_i$, the 
spin-isospin recoupling coefficients 
$<S_iT_i|S_jT_j>_{ST}$, and the 
orbital angular momentum recoupling coefficients 
$A_L^{\ell_i\lambda_i\ell_j\lambda_j}(p_i^\prime q_i p_j q_j)$, 
including the momentum variables $p_i^\prime$ and $p_j$, are 
defined by Eqs. (8)$-$(13) of Ref. \cite{Val05b}.

The integral equations (\ref{e22c5}) couple the amplitude $\psi_i$ to the
amplitudes $\psi_j$ and $\psi_k$. In the cases when the three 
particles are identical or when two are identical and one is different
the equations can be reduced to integral equations
involving just one of the three amplitudes as it has been shown in Ref. 
\cite{Val05b}. In 
contrast, when the three particles are different,
by substituting the equation for 
$\psi_i$ into the corresponding equations for $\psi_j$ and $\psi_k$,
one obtains at best integral equations that
involve two independent amplitudes which
means that in that case the 
numerical calculations are more time consuming.

We will now describe the application of the Faddeev method to
the case of three distinguishable particles.
If one represents in Eq. (\ref{e22c5}) the integration over $dq_j$ by a
numerical quadrature \cite{Abr72}, then for a given set of the 
conserved quantum numbers $L$, $S$, and $T$, 
Eq. (\ref{e22c5}) can be written in the 
matrix form 

\begin{equation}
\psi_i = \sum_{j\ne i} B_{ij}(E) \psi_j,
\label{mex1}
\end{equation}
where $\psi_i$ is a vector whose elements correspond to the values
of the indices $n$, $\ell_i$, $\lambda_i$, $S_i$, $T_i$, and $r$,
i.e., 

\begin{equation}
\psi_i \equiv \psi_{i;LST}^{n\ell_i\lambda_i S_i T_i}(q_r),
\label{mex2}
\end{equation}
with $q_r$ the abscisas of the integration quadrature.
The matrix $B_{ij}(E)$ is given by

\begin{equation}
B_{ij}(E) \equiv q_s^2w_sK_{ij;LST}^{n\ell_i\lambda_i S_i T_i
m\ell_j\lambda_jS_jT_j}(q_r,q_s;E),
\label{mex3}
\end{equation} 
where the vertical direction
is defined by the values
of the indices $n$, $\ell_i$, $\lambda_i$, $S_i$, $T_i$, and $r$
while the horizontal direction
is defined by the values
of the indices $m$, $\ell_j$, $\lambda_j$, $S_j$, $T_j$, and $s$.
$q_s$ and $w_s$ are the abscisas and weights of the integration
quadrature.

Substituting the Eq. (\ref{mex1}) for $i=1$ into the corresponding
equations for $i=2$ and $i=3$ one obtains

\begin{equation}
[B_{21}(E)B_{12}(E)-1]\psi_2+[B_{21}(E)B_{13}(E)+B_{23}(E)]\psi_3=0,
\label{mex4}
\end{equation}

\begin{equation}
[B_{31}(E)B_{12}(E)+B_{32}(E)]\psi_2+
[B_{31}(E)B_{13}(E)-1]\psi_3=0,
\label{mex5}
\end{equation}
so that the binding energies of the system are the zeroes of
the Fredholm determinant

\begin{equation}
|M(E)|=0,
\label{mex6}
\end{equation}
where

\begin{equation}
M(E)=\pmatrix{B_{21}(E)B_{12}(E)-1 & B_{21}(E)B_{13}(E)+B_{23}(E) \cr
B_{31}(E)B_{12}(E)+B_{32}(E)  & B_{31}(E)B_{13}(E)-1 \cr}.
\label{mex7}
\end{equation}

\section{Results and discussion}
\label{sec4}

The results we are going to present have been obtained by solving
exactly the Schr\"odinger equation by the Faddeev method in momentum
space as we have just described. 

Our results are shown in Tables \ref{t3} 
and \ref{t4} for charmed and beauty baryons, respectively. 
As we are asking for consistency
with the light baryon spectra, let us note that the corresponding results for light
baryons are given in Fig. 2 of Ref. \cite{Val05b}.
Comparing the gross structure of the known experimental states and the
theoretical spectrum we find a good overall agreement.
One can observe that, in general, the supposed orbital excitations (negative
parity states) are predicted lower than experiment,
we will return to this point at the end of this 
section. Note that the radial excitation of $1/2^+$ baryons
is always around 350 MeV above the ground state, both for
charmed and beauty baryons. The only exception is the
$\Xi_i(1/2^+)$ ($i=c$ or $b$) with an excitation energy around
80 MeV. This resonance is not indeed a radial excitation.
The ground state corresponds to a $us$ pair in a singlet 
spin state while the excited one corresponds to the same
pair in a triplet spin state. These two levels are often
denoted in the literature as $\Xi_i(1/2^+)$ and
$\Xi'_i(1/2^+)$. 

Let us first stress the flavor independence of the confining
interaction. It was fixed in Ref. \cite{Val05b}
to drive the nucleon Roper resonance to its correct position
and this is the value, denoted as (A) in Table \ref{t2},
we will use all along this work except for the results shown 
in Fig. \ref{fig4}. Once 
confinement is fixed in the nucleon sector it provides with the
needed strength for baryons of different flavor. 

Another interesting
feature arising from our results is the clear flavor dependence of 
the regularization of the delta function appearing in the OGE. In the first 
calculations of the heavy baryon spectra this regularization 
was done by calculating the $\delta-$function perturbatively
in a properly cut Hilbert space \cite{Cop79}. When dealing with exact solutions,
as those based on Faddeev equations, the delta function has to be
regularized. This regularization has to be flavor dependent if 
one wants to obtain the experimentally observed spin splitting,
i.e., mass difference between the ground $3/2^+$ and $1/2^+$ states.
In the case of pairs of quarks of similar mass one can use 
for the regularization parameter $r_0$
a formula depending on the reduced mass of the system
as has been proved in the past \cite{Sil96,Vij04,Val05b}. However, in the
case of heavy baryons the masses of the interacting quarks 
may be quite different, such that
a reduced mass based formula will give the same result for a
light-charm than for a light-bottom pair. As the color-magnetic
term of the one-gluon exchange interaction depends on the inverse of
the product of the masses of the interacting quarks, such a
potential will be strongly reduced for the heavier pair
producing a too small spin-splitting.
This is the reason why Refs. \cite{Sil96,Alb03} 
predict a small spin-splitting for 
beauty baryons (see the AL1 results in Table \ref{t5}),
because the parameters were adjusted on
the charm sector and a reduced mass based formula was used
to move to beauty baryons. 

The interplay between the pseudoscalar
and the one-gluon exchange interactions is a key problem 
for both the baryon spectra and the two-nucleon system \cite{Nak00}. 
This can be illustrated noting that while 
the $\Sigma_i(3/2^+)-\Lambda_i(1/2^+)$ 
mass difference varies slowly from the strange to the bottom sector,
the $\Sigma_i(3/2^+)-\Sigma_i(1/2^+)$ mass difference varies very fast
(see Table \ref{t5}). The former mass difference
is given by the pseudoscalar and one-gluon exchange forces (see
columns $V_1$ and $V_3$ of Table III in Ref. \cite{Val05b}), 
but the last one is only provided by the one-gluon exchange interaction between
the light and the heavy or strange quark. This is very easy
to understand when the wave function is explicitly constructed. In
the case of $\Lambda$ baryons the two light quarks are in a 
flavor antisymmetric spin $0$
state, the pseudoscalar and the one-gluon exchange forces being
both attractive. For $\Sigma$ baryons they are in a flavor
symmetric spin $1$ state. The pseudoscalar force, being still
attractive, is suppressed by one order of magnitude due to the
expectation value of the $(\vec \sigma_i \cdot \vec \sigma_j)
(\vec \lambda_i \cdot \vec \lambda_j)$ operator \cite{Val05b}
and the one-gluon exchange between the two light quarks
becomes repulsive. Therefore, the attraction is provided by
the interaction between the light and the heavy or strange
quarks, which for heavy quarks $c$ or $b$ is given only
by the one-gluon exchange potential. 
In Table \ref{t7} we have calculated
the mass of $\Sigma_i$ and $\Lambda_i$ ($i=s$ or $b$) 
baryons with and without the
pseudoscalar contribution. As can be seen the effect of the 
pseudoscalar interaction between the two-light quarks is 
approximately the same independently of the third quark.
As the mass difference between the $\Sigma_i(3/2^+)$ and
$\Sigma_i(1/2^+)$ states decreases when increasing the mass of the baryon,
being almost constant the effect of the one pion-exchange, the 
remaining mass difference has to be
accounted for by the one-gluon exchange. 
When using an ad hoc combination
of gluon and boson exchanges one may generate a too large
$\Sigma_i(3/2^+)-\Lambda_i(1/2^+)$ ($i=c$ or $b$) mass difference
and a too small $\Sigma_i(3/2^+)-\Sigma_i(1/2^+)$ mass
difference (see the AL1$_\chi$ results in Table \ref{t5}).
This rules out any ad hoc recipe for the relative strength of both
potentials, in any manner consistent with experiment.
An incorrect scaling of the regularization parameter or a wrong balance
between pseudoscalar and one-gluon exchange contributions
drive incorrect results. This reinforces the importance
of constraining models for the baryon spectra in the 
widest possible set of experimental data. 

\begin{figure}[tbp]
\caption{(a) $\Xi_b(1/2^+)[nsb]$, (b) $\Xi_c(1/2^+)[nsc]$, and (c) $\Xi(1/2^+)[nss]$
ground state masses as a function of the regularization parameter of the
light-strange subsystem, $r_0^{ns}$.}
\label{fig2}
\mbox{\epsfxsize=140mm\epsffile{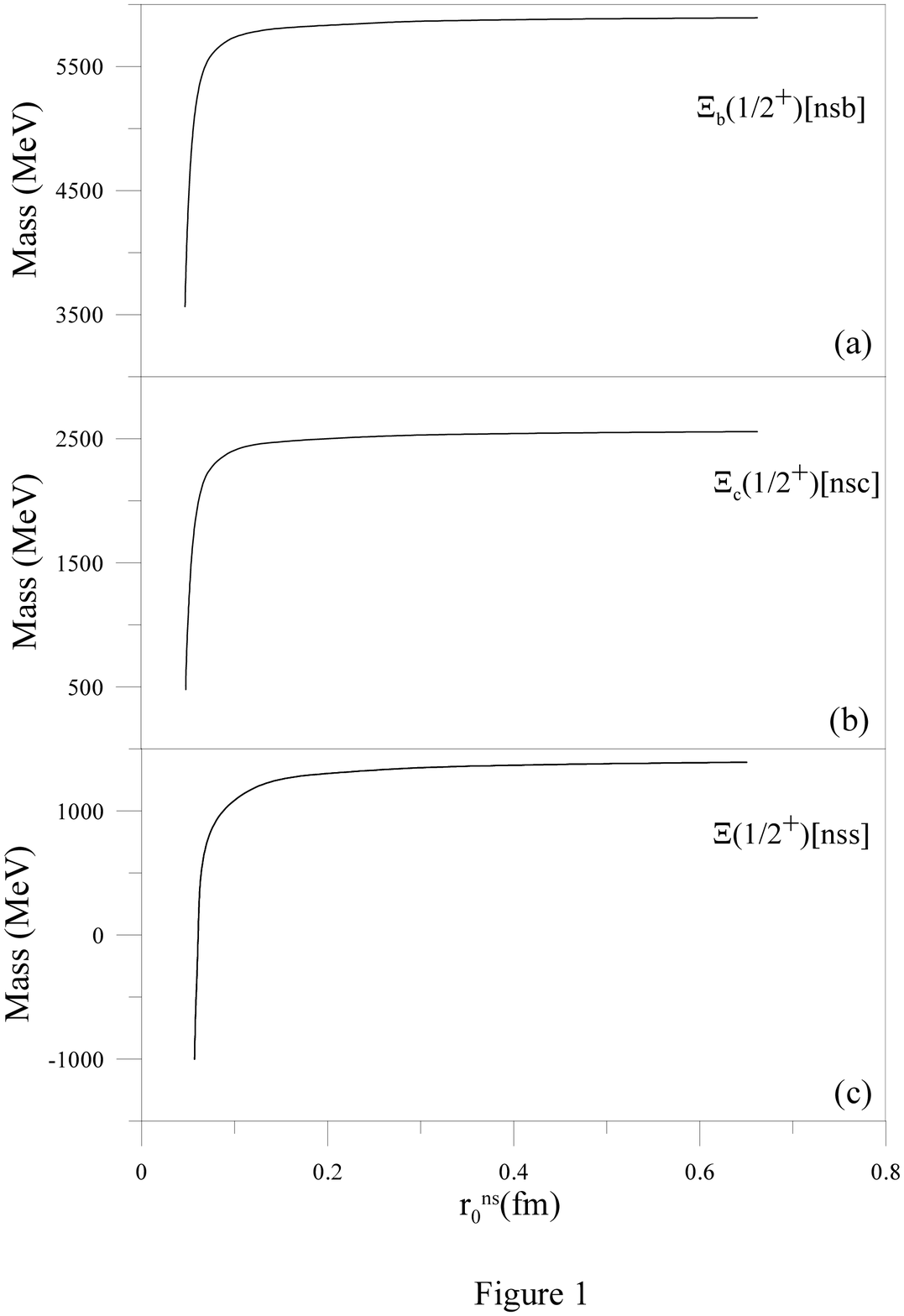}}
\end{figure}

\begin{figure}[tbp]
\caption{(a) $\Xi_b(1/2^+)[nsb]$ ground state mass as a function of the 
regularization parameters $r_0^{nb}$ and $r_0^{sb}$.
(b) $\Xi_c(1/2^+)[nsc]$ ground state mass as a function of the 
regularization parameters $r_0^{nc}$ and $r_0^{sc}$.}
\label{fig3}
\mbox{\epsfxsize=140mm\epsffile{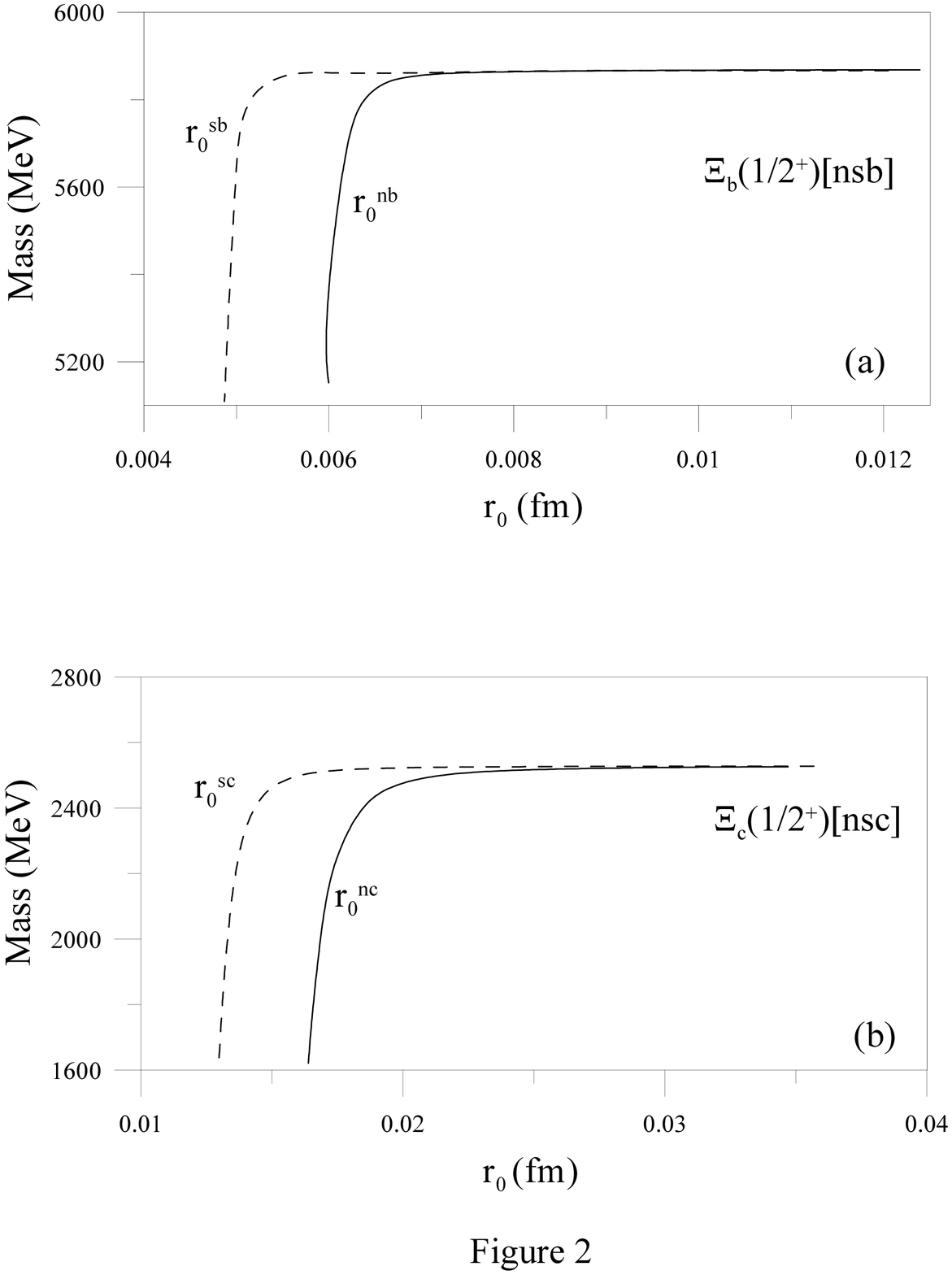}}
\end{figure}

\begin{figure}[tbp]
\caption{(a) Spectra of $\Lambda_c$ for two different confinement
strengths compared to experiment. (b) Same as (a) for $\Sigma_c$ states.}
\label{fig4}
\mbox{\epsfxsize=140mm\epsffile{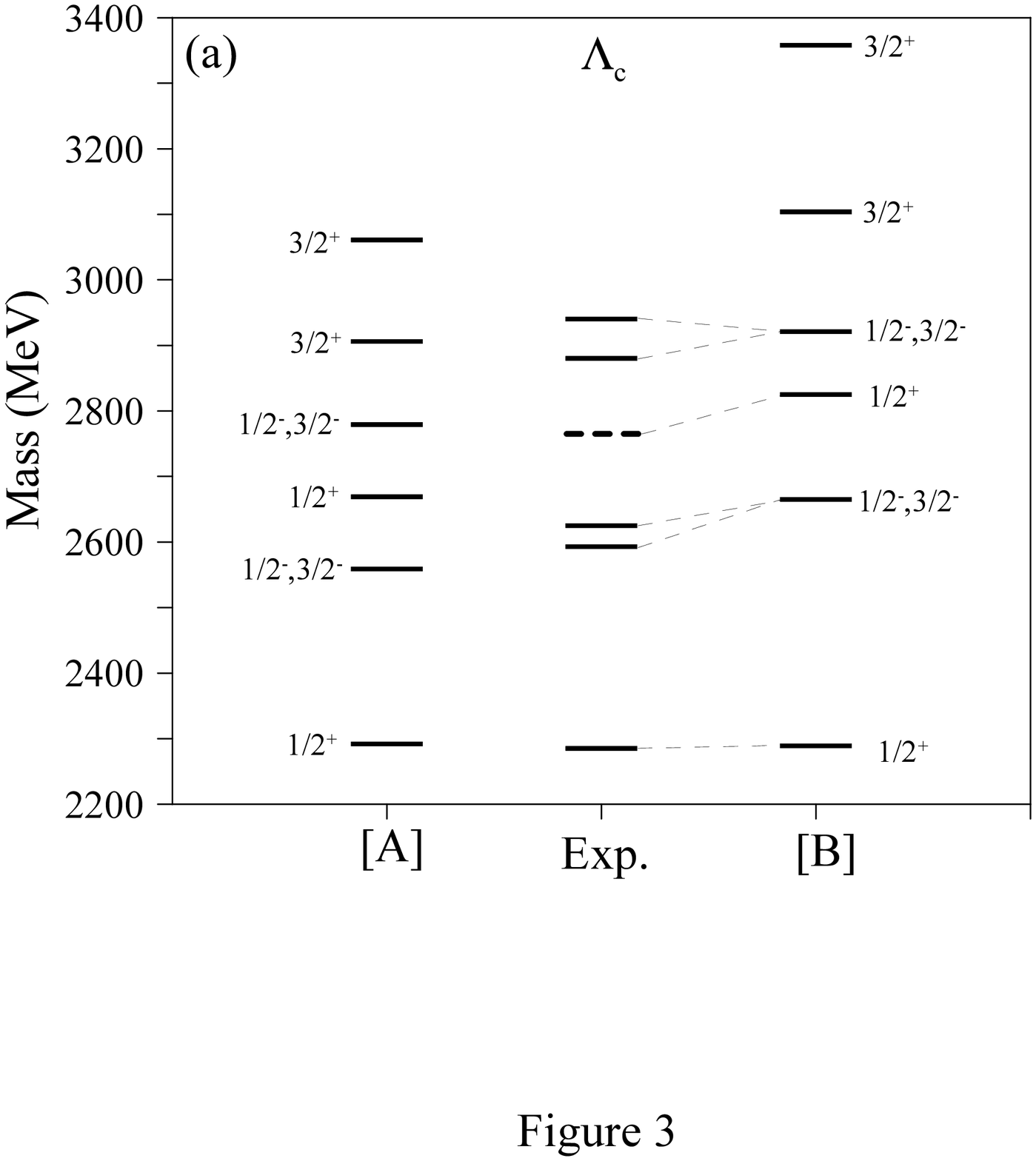}}
\newpage
\mbox{\epsfxsize=140mm\epsffile{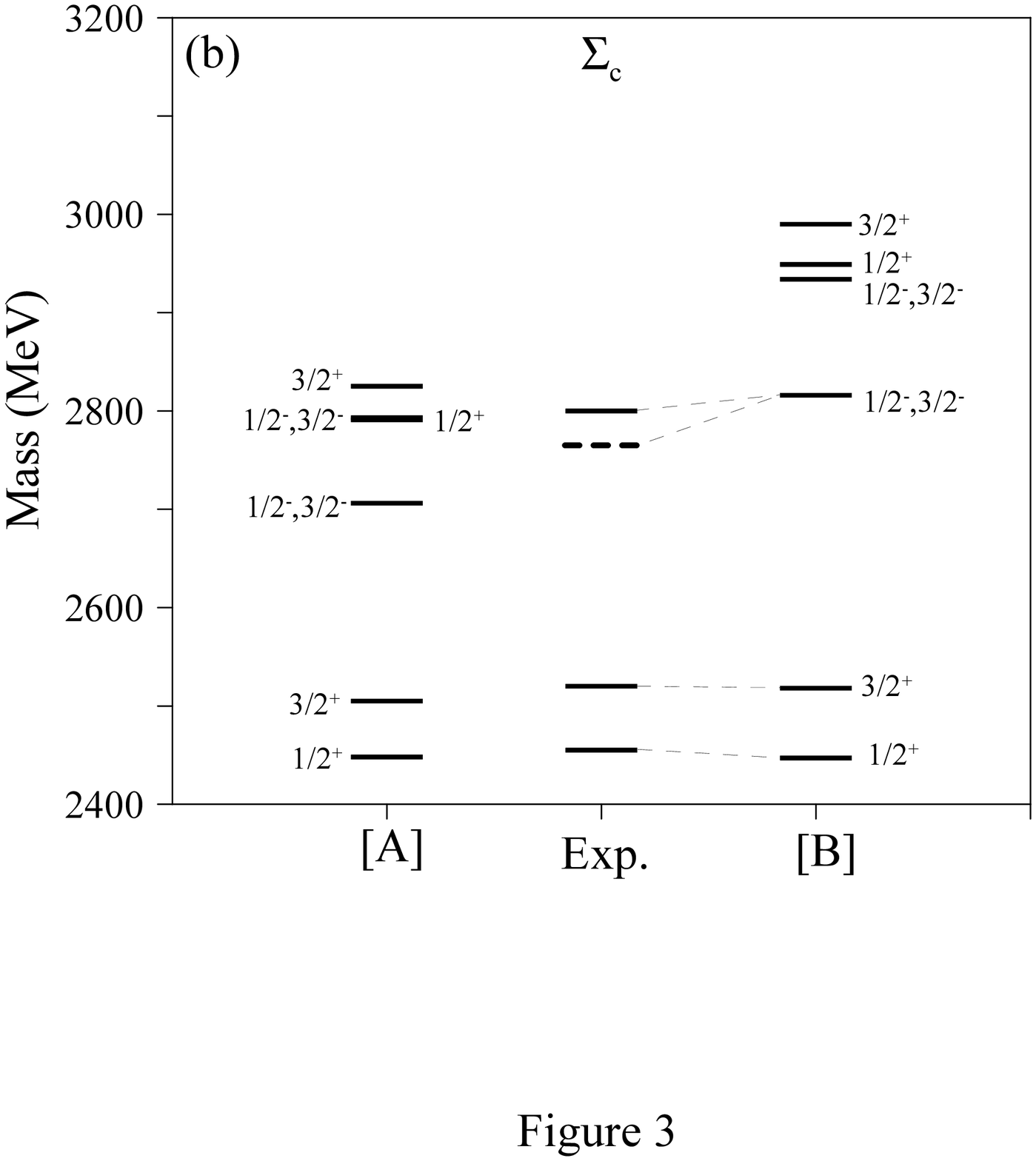}}
\end{figure}

Let us therefore face the problem of the 
regularization parameter of the OGE, $r_0$. As mentioned
in Sec. \ref{sec2} this parameter must vary with the
masses of the interacting quarks. The larger the system (the lighter the masses
of the quarks involved) the larger the value of $r_0$ that can be used without risk
of collapse. In Fig. \ref{fig2} we plot the mass of the $\Xi(1/2^+)[nss]$,
$\Xi_c(1/2^+)[nsc]$ and $\Xi_b(1/2^+)[nsb]$ ground states as a function of the 
regularization parameter of the one-gluon exchange potential for the
$ns$ subsystem, $r_0^{ns}$, for a fixed value of the other two,
$r_0^{nb}$ and $r_0^{sb}$. We observe how the variation of the energy
is very similar in all three cases.
It seems that the dynamics of any two-particle subsystem is not affected
by the nature of the third particle. In particular, the unstable area for 
the one-gluon exchange interaction starts exactly at the same value. 
It is one of the Faddeev amplitudes that becomes unstable independently
of the other two. Therefore the regularization 
parameter should depend on the interacting pair, independently of the
baryon the pair belongs to. The values of $r_0$ reproducing
the experimental data are quoted in Table \ref{t6}. They obey a formula
depending on the product of the masses of the interacting quarks that can
be represented by
\begin{equation}
r_0^{q_iq_j}= A \mu \left( m_{q_i} m_{q_j} \right)^{-3/2}
\end{equation}
being $A$ a constant and $\mu$ the reduced mass of the interacting quarks.
While working with almost equal or not much different masses this law can be
easily replaced by a formula depending on some inverse power of the mass (or
reduced mass) of the pair as obtained in Ref. \cite{Vij04}, but this is not any more the
case for quarks with very different masses, like those present
in heavy baryons. This is one of the reasons why these systems constitute
an excellent laboratory for testing low-energy QCD realizations.
For the sake of completeness we have plotted in 
Fig. \ref{fig3} the mass of the $\Xi_c(1/2^+)$ 
and $\Xi_b(1/2^+)$ states as a function of the $r_0$ values of the two 
subsystems with a heavy and a light quark. We observe
how the value where the unstable region starts is greatly reduced
in going from the $nc$ ($sc$) to the $nb$ ($sb$) subsystem, what
should rule out a reduced mass dependent law that would imply 
a similar value of $r_0$ in both cases.

This behavior of the regularization parameter for large mass
differences of the interacting quarks was not known, and therefore
not taken into account
in Ref. \cite{Vij04b}, predicting a very small 
$3/2^+ -1/2^+$ spin splitting for 
double charmed baryons. This is exactly the same 
problem observed in the calculations of Refs. \cite{Sil96}
and \cite{Alb03}. When the calculation of
Ref. \cite{Vij04b} is repeated with the scaling law of the 
regularization parameter
derived in this work, the spin splitting augments up to
\begin{eqnarray}
\Xi_{cc}(3/2^+)-\Xi_{cc}(1/2^+) &=& 66 \,\, {\rm MeV} \cr
\Omega_{cc}(3/2^+)-\Omega_{cc}(1/2^+) &=& 54 \,\, {\rm MeV} \, .
\end{eqnarray}
Apart from the spin-splitting, the structure 
of the spectra is not significantly modified. In the case
of the triply charmed baryons the results would remain the same
as in Ref. \cite{Vij04b}. 

A widely discussed issue on the light baryon spectra has been
the so-called level ordering problem, the experimentally opposite order
of the $N^*(1440)$ $J^P=1/2^+$ and the $N^*(1535)$ $J^P=1/2^-$ compared
to the harmonic limit. Theoretically, this situation in the light
baryon spectra has been cured by means of appropriate phenomenological
interactions as it is the case of anharmonic terms \cite{Isg00},
scalar three-body forces \cite{Des92}, or pseudoscalar interactions
\cite{Glo96,Gar02}.
For heavy baryons these mechanisms are not expected to work or their
strength is strongly diminished in such a way that the level ordering problem
may not be present. In the case of the
scalar three-body force of Ref. \cite{Des92},
a simultaneous exchange of a scalar particle
among the three quarks, if existing in nature,
should not be active in the presence of a heavy quark due to the
explicit breaking of chiral symmetry.
In the case of the chiral pseudoscalar interaction, its
$(\vec{\sigma}_i \cdot \vec{\sigma}_j) (\vec{\lambda}_i \cdot \vec{\lambda}_j)$
structure gives attraction for symmetric spin-flavor pairs and
repulsion for antisymmetric ones.
The reduction of the strength of the pseudoscalar potential,
acting only between the two light quarks, makes this level
ordering inversion highly not probable for heavy baryons.
This effect was already noticed in the strange
baryon spectra. We observed how the one pion exchange
is very much reduced in the
case of $\Sigma(1/2^+)$ states (see Table III of Ref. \cite{Val05b}),
its role being replaced by the kaon exchange between the light
and {\it heavy} (strange in that case) quarks. Even then,
the strength is not enough to reverse the ordering of the states
(see Fig. 3 of Ref. \cite{Val05b}). When the $SU(3)$ symmetry is broken
and one puts a heavy quark ($c$ or $b$) the situation is clearly
favored for having a first negative parity excited state below
the Roper resonance. The presence of the heavy quark
will also diminish the importance of relativistic effects,
being responsible for most part of the level ordering problem \cite{Gar03}.
We consider that the determination of the position
of the first radial excitation of the $\Sigma_c$ could
help in understanding this problem.

Our results allow to predict quantum numbers to the
experimentally observed resonances. These assignments are not mandatory.
At this point we have to return to the observation that
the orbital excitations are predicted lower than observed.
This is a consequence of fitting the confinement strength 
in the light baryon sector such as to reproduce the nucleon Roper 
resonance instead of the negative parity states. If we modify
the confinement strength to reproduce the negative parity 
states in the light baryon sector (this would only imply to
loose the description of the nucleon Roper resonance that it is
known to be highly sensitive to relativistic kinematics \cite{Gar03})
and recalculate the spectra with the new confining strength, denoted 
by (B) in Table \ref{t2}, we obtain the results shown in 
Fig. \ref{fig4}. 
Let us notice that all our previous
discussion and the conclusions derived hold for the new confinement strength.
In Fig. \ref{fig4}, column [A] represents the results of the
smaller confinement strength and [B] those of the larger
confinement strength in Table \ref{t2}. A much better agreement is observed
with the model reproducing the orbital excitations of the
light baryon sector. There is no experimental
state that we do not predict and there is no low-lying
theoretical resonance that has not been observed.
The smaller confinement strength would give rise
to several states still not observed. The recently discovered
$\Sigma_c(2800)$ \cite{Miz05} would correspond to an
orbital excitation with $J^P=1/2^-$ or $3/2^-$ (they are
degenerate in our model), any other correspondence being
definitively excluded. For $\Lambda_c$ baryons, the recently
confirmed as a $\Lambda_c$ state, $\Lambda_c(2880)$ \cite{Aub06},
and the new state $\Lambda_c(2940)$ \cite{Aub06} may constitute
the second orbital excitation of the $\Lambda_c$ baryon. Finally,
there is an state with a mass of 2765 MeV reported in Ref. \cite{Art01}
as a possible $\Lambda_c$ or $\Sigma_c$ state and also observed
in Ref. \cite{Miz05}. While the first reference (and also the PDG)
are not able to decide between a $\Lambda_c$ or a $\Sigma_c$
state, the second one prefers a $\Lambda_c$ assignment.
As seen in Fig. \ref{fig4}, 
this state may constitute the second member of the
first orbital excitation of $\Sigma_c$ states or the first radial
excitation of $\Lambda_c$ baryons. An experimental effort to
confirm the existence of this state and its decay modes would
help on the symbiotic process between experiment and theory
to disentangle the details of the structure of heavy baryons.

Let us finally note that heavy quark symmetry (HQS) and chiral symmetry
can be combined together in order to describe the soft hadronic
interactions of hadrons containing a heavy quark \cite{Wis92}. In the
limit of the heavy quark mass $m_Q \to \infty$ HQS predicts that all
states in the $\bf 6$ $SU(3)$ representation (those where the light 
degrees of freedom are in a $s=1$ state) would be degenerate.
If one considers 
HQS and lowest order $SU(3)$ breaking \cite{Sav95} one obtains an equal spacing rule
similar to the one that arises in the decuplet of uncharmed $J^P=3/2^+$
baryons. The equal spacing rule obtained for charmed baryons is

\begin{eqnarray}
\Xi'_c(1/2^+) - \Sigma_c(1/2^+) &=&
\Omega_c(1/2^+) - \Xi'_c(1/2^+)= \nonumber \\
= \Xi_c(3/2^+) - \Sigma_c(3/2^+) &=&
\Omega_c(3/2^+) - \Xi_c(3/2^+)
\label{eqs}
\end{eqnarray}
These relations are satisfied by experimental data as seen in the second
column of Table \ref{t10}, the first three spacings being of the 
order of 123 MeV. This prediction is clearly sustained by our model
giving rise to a spacing of 127 MeV (third column of Table \ref{t10})
and also by the results of OGE
based models (fourth column of Table \ref{t10}). Lattice calculations
based on HQS fulfill exactly this equal spacing rule (fifth column
of Table \ref{t10}). However, the relativistically covariant quark 
model of Ref. \cite{Mig06} strongly violates this rule, contradicting
the expectations of HQS. All these results allow to confirm the prediction 
of a $\Omega_c(3/2^+)$ state at around 2770 MeV.

This equal spacing rule should also held in the beauty baryon sector, giving rise
to the relations

\begin{eqnarray}
\Xi'_b(1/2^+) - \Sigma_b(1/2^+) &=&
\Omega_b(1/2^+) - \Xi'_b(1/2^+)= \nonumber \\
= \Xi_b(3/2^+) - \Sigma_b(3/2^+) &=&
\Omega_b(3/2^+) - \Xi_b(3/2^+) \, .
\label{eqsb}
\end{eqnarray}
Once again this equal spacing rule is strictly fulfilled by our model, generating
an spacing of 124 MeV in all four cases. If we now make use of the
existing indications of experimental data for $\Sigma_b(1/2^+)$ and
$\Sigma_b(3/2^+)$, then we can predict the existence of the following
states: a $\Xi'_b(1/2^+)$
with a mass of 5920 MeV, a $\Omega_b(1/2^+)$ with a mass of 6044 MeV,
a $\Xi_b(3/2^+)$ with a mass of 5976 MeV, and finally a $\Omega_b(3/2^+)$
with a mass of 6100 MeV.

\section{Summary}
\label{sec5}

We have used a constituent quark model incorporating the basic properties
of QCD to study the heavy baryon spectra. 
Consistency with the baryon-baryon interaction and the light baryon spectra
is asked for. The model takes into account the most
important QCD nonperturbative effects: chiral symmetry breaking
and confinement as dictated by unquenched lattice QCD. It also
considers QCD perturbative effects trough a flavor dependent one-gluon exchange
potential.  

The three-body problem has been exactly solved by means of the
Faddeev method in momentum space. For the first time we have
studied baryons made of three different quarks, what makes
the calculation time consuming, but providing an excellent
test of our numerical method. We have found that the key interplay
between pseudoscalar and one-gluon exchange forces, already observed
for the light baryons, constitutes a basic ingredient for the 
description of heavy baryons.
The final spectra results from a subtle but physically meaningful
balance between different spin-dependent forces. The baryon spectra make 
manifest the presence of two different sources of spin-dependent
forces that can be very well mimic by the operatorial dependence
generated by the pseudoscalar and one-gluon exchange potentials.

While the flavor dependence of the regularized one-gluon exchange potential
for equal mass quarks can be nicely described by the inverse
of the reduced mass of the system, we have found that this is not 
the case for interacting quarks with large mass differences. It
is instead a mass dependence considering explicitly the masses of 
the two quarks that provides with a nice agreement with data.

Heavy baryons constitute an extremely interesting problem
joining the dynamics of light-light and heavy-light subsystems
in an amazing manner. This is due to the remnant effect
of pseudoscalar forces in the two-light quark subsystem. Models based
only on boson exchanges cannot explain in any manner the
dynamics of heavy baryons, but it becomes also difficult for models
based only on gluon exchanges, if consistency between light and
heavy baryons is asked for. One-gluon exchange models would reduce
the problem to a two-body problem controlled by the dynamics of the
heaviest subsystem, and we find evidences in the spectra for 
contributions of both subsystems.

Our results contain the equal mass spacing rules obtained for heavy
baryons by means of heavy quark symmetry and lowest order
$SU(3)$ chiral symmetry breaking. The obtained spacing is almost the
same for charmed and beauty baryons. Making use of the available 
experimental data we can predict the existence of six new 
beauty baryons, their masses and quantum numbers,
as well as we can confirm the prediction of a charmed
baryon made some time ago. We also make predictions
for the orbital and radial excitations of all quantum numbers
that, if confirmed, would definitively prove the flavor
independence of the confining interaction.

Finally, although we do not believe that explanations
based on constituent quark models may rule out or contradict
other alternative ones, one should acknowledge the capability
of constituent quark models 
for a coherent understanding of the low-energy phenomena of the
baryon spectroscopy and the baryon-baryon interaction in a simple 
framework based on the contribution of pseudoscalar, scalar
and one-gluon-exchange forces between quarks.

\section{acknowledgments}

This work has been partially funded by Ministerio 
de Ciencia y Tecnolog{\'{\i}}a under Contract No. FPA2004-05616,
by Junta de Castilla y Le\'{o}n under Contract No. SA-104/04,
and by COFAA-IPN (M\'exico).

\begin{table}[tbp]
\caption{Experimentally known charmed and beauty baryons.
The $J$ and $P$ quantum numbers have not been measured and we have
quoted, when existing, the values given by the PDG that correspond to 
the quark model expectations. We have omitted the very
small error bars in the mass. We have quoted in the last column the quark
content of the different symbols used to denote the baryons, $n$ stands
for a light quark $u$ or $d$. 
We have also included two charmed baryons recently reported in
Refs. \protect\cite{Miz05} and \protect\cite{Aub06}, 
but not appearing for the moment in the PDG. Finally, we also quote
two beauty baryons suggested in Ref. \protect\cite{Fei95}. 
For completeness, in the bottom part of the table we have 
included two double charmed baryons that will appear along the text.}
\label{t1}
\begin{center}
\begin{tabular}{|ccccccc|}
& State & $J^P$ & Mass (MeV) & Status  & Quark content & \\
\hline
& $\Lambda_c$                         & $1/2^+$ & 2285 & $****$ & $udc$ & \\
& $\Lambda_c(2593)$                   & $1/2^-$ & 2594 & $***$  &       & \\
& $\Lambda_c(2625)$                   & $3/2^-$ & 2627 & $***$  &       & \\
& $\Lambda_c$ or $\Sigma_c (2765) $ & $?^?  $ & 2765 & $*$    &       & \\
& $\Lambda_c(2880)$                   & $?^?  $ & 2881 & $**$   &       & \\
& $\Lambda_c(2940)$                   & $?^?  $ & 2940 & Ref. \protect\cite{Aub06} &       & \\
& $\Sigma_c(2455)$                    & $1/2^+$ & 2452 & $****$ & $nnc$ & \\
& $\Sigma_c(2520)$                    & $3/2^+$ & 2518 & $***$  &       & \\
& $\Sigma_c(2800)$                    & $3/2^-$ & 2800 & Ref. \protect\cite{Miz05} &       & \\
& $\Xi_c$                             & $1/2^+$ & 2469 & $***$  & $nsc$ & \\
& $\Xi'_c$                            & $1/2^+$ & 2577 & $***$  &       & \\
& $\Xi_c(2645)$                       & $3/2^+$ & 2646 & $***$  &       & \\
& $\Xi_c(2790)$                       & $1/2^-$ & 2790 & $***$  &       & \\
& $\Xi_c(2815)$                       & $3/2^-$ & 2816 & $***$  &       & \\
& $\Omega_c$                          & $1/2^+$ & 2698 & $***$  & $ssc$ & \\
\hline
& $\Lambda_b$                         & $1/2^+$ & 5625 & $***$  & $udb$ & \\
& $\Sigma_b(5796)$                    & $1/2^+$ & 5796 & Ref. \protect\cite{Fei95} & $nnb$ & \\
& $\Sigma_b(5852)$                    & $3/2^+$ & 5852 & Ref. \protect\cite{Fei95} &       & \\
& $\Xi_b$                             & $1/2^+$ &  ?   & $*$    & $nsb$ & \\
& $\Omega_b$                          &         &      &        & $ssb$ & \\
\hline
\hline
& $\Xi_{cc}$                          & $?^?$   & 3519 & $*$    & $ucc$ & \\ 
& $\Omega_{cc}$                       &         &      &        & $scc$ & \\ 
\end{tabular}
\end{center}
\end{table}

\begin{table}[tbp]
\caption{Quark-model parameters. (A) and (B) stand for two different
confinement strengths and charm quark masses used along this work.}
\label{t2}
\begin{center}
\begin{tabular}{|cc|ccc|}
&&$m_u=m_d$ (MeV) & 313 & \\ 
&Quark masses&$m_s$ (MeV)     & 500 & \\ 
&&$m_c$ (MeV) & (A) 1650 / (B) 1740 & \\ 
&&$m_b$ (MeV) & 5024 & \\ 
\hline
&&$m_{\pi}$ (fm$^{-1}$) & 0.70&\\
&&$m_{\sigma}$ (fm$^{-1}$)& 3.42&\\ 
&&$m_{\eta}$ (fm$^{-1}$)  & 2.77&\\ 
&Boson exchanges&$m_K$ (fm$^{-1}$)       & 2.51&\\ 
&&$\Lambda_{\pi}=\Lambda_{\sigma}$ (fm$^{-1}$) & 4.20 &\\
&&$\Lambda_{\eta}=\Lambda_K$ (fm$^{-1}$) & 5.20&\\ 
&& $g_{ch}^2/(4\pi)$      & 0.54&\\ 
&&$\theta_P(^o)$          & $-$15&\\ 
\hline
&&$a_c$ (MeV)             &(A) 230 / (B) 340&\\
&Confinement&$\mu_c$ (fm$^{-1}$)&0.70&\\ 
\hline
&OGE&$r_0$ (fm)     & see text&\\
\end{tabular}
\end{center}
\end{table}

\begin{table}[tbp]
\caption{Masses, in MeV, for baryons with a charm quark with the set (A)
of parameters compared to experiment. Note that we have included a possible
assignment of the state with a mass of 
2765 MeV as a $\Lambda_c$ or $\Sigma_c$ state.}
\label{t3}
\begin{center}
\begin{tabular}{|c|cc|cc|cc|cc|}
 &\multicolumn{2}{c|}{$J^P=1/2^+$} & \multicolumn{2}{c|}{$J^P=3/2^+$} & 
\multicolumn{2}{c|}{$J^P=1/2^-$} & \multicolumn{2}{c|}{$J^P=3/2^-$}  \\ 
 & Exp. & Theor.  & Exp. & Theor.  & Exp. & Theor.  & Exp. & Theor. \\
\hline
 $\Lambda_c$  & 2285 & 2292 & 2940$\tablenotemark\tablenotetext{Ref. \cite{Aub06}}$ & 2906 & 2593 & 2559 & 2625 & 2559 \\
              & 2765$\tablenotemark\tablenotetext{Ref. \cite{Art01}}$ & 2669 & $-$ & 3061 & 2880${\rm b}$
& 2779 & 2880$^{\rm b}$ & 2779 \\
\hline
 $\Sigma_c$  & 2455 & 2448 & 2520 & 2505 & 2765$^{\rm b}$  & 2706 & 2765$^{\rm b}$  & 2706 \\
             & $-$  & 2793 & $-$  & 2825 & 2800$\tablenotemark\tablenotetext{Ref. \cite{Miz05}}$
& 2791 & 2800$^{\rm c}$ & 2791 \\
\hline
 $\Xi_c$     & 2468 & 2496 & 2645 & 2633 & 2790 & 2749 & 2815 & 2749 \\
             & 2576 & 2574 & $-$  & 2951 & $-$  & 2829 & $-$  & 2829 \\
\hline
 $\Omega_c$  & 2697 & 2701 & $-$  & 2759 & $-$  & 2959 & $-$  & 2959 \\
             & $-$  & 3044 & $-$  & 3080 & $-$  & 3029 & $-$  & 3029 \\
\end{tabular}
\end{center}
\end{table}

\begin{table}[tbp]
\caption{Masses, in MeV, for baryons with a bottom quark compared to experiment.}
\label{t4}
\begin{center}
\begin{tabular}{|c|cc|cc|cc|cc|}
 &\multicolumn{2}{c|}{$J^P=1/2^+$} & \multicolumn{2}{c|}{$J^P=3/2^+$} & 
\multicolumn{2}{c|}{$J^P=1/2^-$} & \multicolumn{2}{c|}{$J^P=3/2^-$}  \\ 
 & Exp. & Theor.  & Exp. & Theor.  & Exp. & Theor.  & Exp. & Theor. \\
\hline
 $\Lambda_b$ & 5624 & 5624 & $-$ & 6246 & $-$ & 5890 & $-$ & 5890 \\
             & $-$  & 5996 & $-$ & 6406 & $-$ & 6132 & $-$ & 6132 \\
\hline
 $\Sigma_b$  & 5796$\tablenotemark\tablenotetext{Ref. \cite{Fei95}}$
& 5789 & 5852$^{\rm a}$ & 5844 & $-$  & 6039 & $-$  & 6039 \\
             & $-$  & 6127 & $-$  & 6158 & $-$ & 6142 & $-$ & 6142 \\
\hline
 $\Xi_b$     & $-$  & 5825 & $-$  & 5967 & $-$  & 6076 & $-$  & 6076 \\
             & $-$  & 5913 & $-$  & 6275 & $-$  & 6157 & $-$  & 6157  \\
\hline
 $\Omega_b$  & $-$  & 6037 & $-$  & 6090 & $-$  & 6278 & $-$  & 6278 \\
             & $-$  & 6367 & $-$  & 6398 & $-$  & 6373 & $-$  & 6373 \\
\end{tabular}
\end{center}
\end{table}

\begin{table}[tbp]
\caption{Mass difference (in MeV) between $\Sigma_i$ and $\Lambda_i$ states for 
different flavor sectors.}
\label{t5}
\begin{center}
\begin{tabular}{|ccccc|}
Mass difference  & Exp.  & This work &  AL1$_\chi$ \protect\cite{Alb03} & AL1\protect\cite{Sil96} \\
\hline
$\Sigma(3/2^+)-\Lambda(1/2^+)$     & 269   &    260    &   $-$       & $-$ \\
$\Sigma(3/2^+)-\Sigma(1/2^+)$      & 195   &    169    &   $-$       & $-$ \\
$\Sigma_c(3/2^+)-\Lambda_c(1/2^+)$ & 235   &    223    &   395       & 253 \\
$\Sigma_c(3/2^+)-\Sigma_c(1/2^+)$  &  65   &     57    &    78       &  79 \\
$\Sigma_b(3/2^+)-\Lambda_b(1/2^+)$ & 228   &    220    &   394       & 239 \\
$\Sigma_b(3/2^+)-\Sigma_b(1/2^+)$  &  56   &     55    &    28       &  31 \\
\end{tabular}
\end{center}
\end{table}

\begin{table}[tbp]
\caption{Masses, in MeV, of different beauty baryons with two-light
quarks with (Full) and without ($V_\pi=0$) the contribution of the one-pion exchange
potential. The same results have been extracted from Table III of 
Ref. \protect\cite{Val05b} for strange baryons. $\Delta E$ stands 
for the difference between both results.}
\label{t7}
\begin{center}
\begin{tabular}{|cccccc|}
& State & Full & $V_\pi=0$ & $\Delta E$ & \\
\hline
& $\Sigma_b(1/2^+)$     & 5789 & 5802 & $-$13  & \\
& $\Sigma_b(3/2^+)$     & 5844 & 5854 & $-$10  & \\
& $\Lambda_b(1/2^+)$    & 5624 & 5804 & $-$180  & \\
& $\Lambda_b(3/2^+)$    & 6246 & 6246 & $<$ 1  & \\
\hline\hline
& State & $V_{CON}+V_{OGE}+V_\pi$ & $V_{CON}+V_{OGE}$ & $\Delta E$ & \\
\hline
& $\Sigma(1/2^+)$     & 1408 & 1417 & $-$9  & \\
& $\Sigma(3/2^+)$     & 1454 & 1462 & $-$8  & \\
& $\Lambda(1/2^+)$    & 1225 & 1405 & $-$180 & \\
\end{tabular}
\end{center}
\end{table}

\begin{table}[tbp]
\caption{$r_0^{q_iq_j}$ in fm.}
\label{t6}
\begin{center}
\begin{tabular}{|cccc|}
& $(q_i,q_j)$ & $r_0^{q_iq_j}$ & \\
\hline
& $(n,n)$     & 0.35 & \\
& $(s,n)$     & 0.2845 & \\
& $(n,c)$     & 0.0347 & \\
& $(s,c)$     & 0.0285 & \\
& $(n,b)$     & 0.0074 & \\
& $(s,b)$     & 0.0060 & \\
\end{tabular}
\end{center}
\end{table}

\begin{table}[tbp]
\caption{Equal spacing rule of Eq. (\ref{eqs}) for different
models in the literature. Masses are in MeV.}
\label{t10}
\begin{center}
\begin{tabular}{|cccccc|}
Mass difference  & Exp.  & This work &  Ref. \protect\cite{Sil96} & 
Ref. \protect\cite{Mig06} & Ref. \protect\cite{Bow96} \\
\hline
$\Xi'_c(1/2^+)-\Sigma_c(1/2^+)$  & 121   &  126  & 112  & 136 & 110 \\
$\Omega_c(1/2^+)-\Xi'_c(1/2^+)$  & 121   &  127  & 108  &  93 & 110 \\
$\Xi_c(3/2^+)-\Sigma_c(3/2^+)$   & 125   &  128  & 112  & 112 & 110 \\
$\Omega_c(3/2^+)-\Xi_c(3/2^+)$   & $-$   &  126  & 103  &  70 & 110 \\
\end{tabular}
\end{center}
\end{table}

\end{document}